\documentclass[iop]{emulateapj}

\newcommand{\HII}{H\,{\sc ii}}

\newcommand{\um}{\,$\mu$m}
\newcommand{\kms}{\,km\,s$^{-1}$}
\newcommand{\wno}{\,cm$^{-1}$}

\slugcomment{}

\shorttitle{Double RISN in II Zw 40}
\shortauthors{Beck \% et al}

\begin{document}

\title{Ionized Gas Kinematics at High Resolution. II: Discovery of a Double Infrared Cluster in II Zw 40  }

\author{Sara Beck\altaffilmark{1,2}, Jean Turner \altaffilmark{3}, John Lacy\altaffilmark{4}, Thomas Greathouse \altaffilmark{2,5} Ohr Lahad \altaffilmark{1}}
\affil{School of Physics and Astronomy, Tel Aviv University , Ramat Aviv, Israel, 69978}

\altaffiltext{1}{School of Physics and Astronomy, Tel Aviv University, Ramat Aviv ISRAEL 69978}
\altaffiltext{2}{Visiting Astronomer at the Infrared Telescope Facility, which is operated by the University of Hawaii under Cooperative Agreement no. NNX-08AE38A with the National Aeronautics and Space Administration, Science Mission Directorate, Planetary Astronomy Program.}
\altaffiltext{3}{Department of Physics and Astronomy, UCLA, Los Angeles, CA 90095-1547}
\altaffiltext{4}{Department of Astronomy, University of Texas at Austin, Austin Tx 78712}
\altaffiltext{5}{Southwest Research Institute, San Antonio Tx 78228-0510}

\begin{abstract}
 The nearby  dwarf galaxy II Zw 40 hosts an intense starburst.  At the center of the starburst is a bright compact radio and infrared source,  thought to be a giant dense \HII~region containing  $\approx14,000$ O stars. Radio continuum images suggest that the compact source is actually a collection of several smaller emission regions.  We accordingly use 
 the kinematics of the  ionized gas to probe the structure of the radio-infrared emission region.   With TEXES on the NASA-IRTF we measured the 10.5\um\ [SIV] emission line with effective spectral resolutions, including thermal broadening, of $\sim25$ and $\sim3$\kms ~and spatial resolution $\sim1$\arcsec.  The line profile shows two distinct, spatially coextensive, emission features. The stronger feature is at galactic velocity and has FWHM 47\kms.  The second feature is $\sim44$\kms~ redward of the first and has FWHM 32\kms.   We argue that these are two giant embedded clusters, and estimate their masses to be $\approx3\times10^5 M_\odot$ and $\approx1.5\times10^5 M_\odot$.  The velocity shift  is unexpectedly large for such a small spatial offset. We suggest that it may arise in a previously undetected kinematic feature remaining from the violent merger that formed the galaxy.  
\end{abstract}


\section{Introduction}

The most intense star formation sites in nearby starbursts  contain hundreds or thousands of OB stars in volumes no more than a few parsec across, embedded in dense obscuring material.   
The stars excite regions of high infrared and radio luminosity and large ionizing fluxes, which are observed as `radio-infrared supernebulae' (RISN). The massive stars will also provide considerable kinetic energy via winds and violent episodes.   
These intense and compact energy sources 
are clearly of great importance for the development of the starburst, and of the entire host galaxy. How will they evolve and how do they affect their surroundings?    To answer those questions we need to understand the kinematics of the gas ionized by the embedded stars.  The extinction to these star formation regions is too high for  optical or ultra-violet tracers to be seen.

This is the third 
in a series on high resolution spectroscopy of intense extragalactic star formation sources in the middle infrared \citep{2010ApJ...722.1175B,BLT12}.   We use fine-structure lines of metal ions  as kinematic tracers, rather than the more commonly observed hydrogen recombination lines, because the metal lines are less affected by thermal broadening. In a nebula with electron temperature $T_e$, the thermal motions effectively convolve lines emitted by an ion of mass $m_i$ with a gaussian of $\sigma=(kT_e/m_i)^{1/2}$ or FWHM $v=2.35\times(kT_e/m_i)^{1/2}$.  For $T_e=10^4$,  the thermal widening of a hydrogen line is $\sim21$\kms.   Heavier elements suffer less thermal broadening  by a factor of $V_m/V_h  = (m_h/m_m)^{1/2}$ and can more accurately detect fine kinematic structure.    The value of metal emission lines as  kinematic probes of
\HII\ regions has been demonstrated for Galactic ultra-compact \HII ~regions \citep{ZL08,ZH,DJ03}, and in \citet{BLT12} for the supernebula in N5253. 

 Here we address the starburst in the nearby dwarf galaxy II Zw 40. II Zw 40 appears to have formed in a  collision or merger between two smaller dwarf galaxies. HI observations  \citep{BK88,VZ98} find two kinematic systems, consistent with the interactions of two HI clouds of $\sim 2\times10^8M_{\odot}$ each.  The starburst is identified with one of the clouds and the other is thought to be quiescent.   \citet{BT02} found a radio and infrared source at $\alpha=5^h55^m42^s.69, \delta=03^o23'32.04\arcsec$ to be the site of the youngest and most intense star formation and  the dominant star formation source of the galaxy.   II~Zw 40  has low metallicity [O]/[H]=8.25 \citep[][]{WR93}, and 
little molecular gas \citep{SSLH92,2001AJ....121..740M}.  
With its huge population of young stars, this implies an 
unusually high star formation 
efficiency  \citep{SSLH92}.  
How is the star formation manifested in this unusually efficient starburst? 
Is it in a single large cluster or extended? Do the emission lines from the
ionized gas indicate how the star formation is affecting its surroundings?

Kinematics of the ionized gas in II Zw 40 have been extensively studied in the optical \HII ~region  \citep[][and references therein]{BPT09}, but there is much less spectroscopic information on the embedded sources, which are observable only in the infrared and radio.  The extinction to the infrared source in II Zw 40 as found from the near-infrared HI lines and the free-free radio flux \citep{BT02} is $A_v\approx 8-10 ~mag$,  over the inner 3\arcsec.  
Near-infrared spectra were obtained by \citet{VA08} and Spitzer IRS mid-infrared spectra by \citet{WU08} but  these studies lacked the 
spectral resolution 
to resolve the emission lines. 
 The infrared observations with the highest spectra resolution to date are $Br\alpha$ and $Br\gamma$ at 4.05\um~and 2.17\um, obtained with NIRSPEC on the Keck Telescope by \citet{TB01}.  Those found  lines of FWHM $\sim100$\kms ~and no obvious non-Gaussian structure.  But while the spectral resolution of NIRSPEC for the Brackett lines is $\sim25,000$ or 12\kms, the thermal broadening  of the hydrogen lines (see previous section) degrades the effective resolution of those observations to  only 24~\kms.  To obtain higher resolution, we must observe  lines emitted by heavy metals. 

II Zw 40 has a  high excitation spectrum and is dominated in the infrared by [NeIII] 15.5\um,  which cannot be observed from the ground \citep{TH00}.   The Infrared Spectrometer on Spitzer recorded the II Zw 40 spectrum with the Short-HI module and found the strongest line that can be observed  from the ground to be [SIV]   951.42\wno ~(10.5\um).  We accordingly observed II Zw 40 in this line.  The observations and data analysis are described in the next section and the results in Section 3.

\section{Observations,  Spatial Distribution and Total Flux of the [SIV] Line}
II Zw 40 was observed at the NASA Infrared Telescope Facility on Mauna Kea, Hawaii, on the nights of 9-10 and 10-11 January 2012, with the TEXES spectrometer \citep{LA02}.  The weather was extremely dry and observing conditions were excellent.  TEXES on the IRTF has a seeing and diffraction-limited beam of $\sim1.4\arcsec$. The observations of 9-10 January were taken in the medium resolution mode.   The slit was  29 pixels long and the plate scale $0.36\arcsec\times8.5~\rm km~s^{-1}$ per pixel.  The telescope was offset pointed to the coordinates of the free-free radio emission peak which \citet{BT02} identified as the core star formation region. Offset pointing  in our experience at the IRTF  is typically accurate to $\sim1~\arcsec$.  The source was nodded N-S along the slit. The line was seen clearly in the first position. Four more spectra were obtained, positioned in steps of $0.7\arcsec$ (half the slit width) east and west. The line was detected in the pointings $0.7\arcsec$ east and west from the first position; it was not detected in the positions further out.  The galaxy and slit are shown in Figure 1 and Figure 2. 

Calibration followed the procedure outlined in \citet{LA02}.  The source spectra are divided by a black-sky flat field, which also provides the radiometic calibration.  Atmospheric lines observed in asteroid spectra give the wavelength calibration. There are weak atmospheric lines at wavelengths equivalent to [SIV] at redshifts $\sim 1000$ and $\sim 600$ \kms~ but the atmosphere at II Zw 40's velocity appeared completely clear.  This is confirmed by the established atmospheric transmission curves for Mauna Kea\footnote{The transmission curves may be found at the Gemini website: http://www.gemini.edu/?q=node/10789.} which show transmission of $.99\pm0.003$ across this wavelength region.  
 
Figure 3 shows the
individual spectra of the three overlapping positions where the line was detected.   The [SIV] emission is highly concentrated;  these observations have only barely resolved it spatially, consistent with the radio continuum appearance (Fig.~2).  %
 There are no convincing variations in line shape  across the source.  The total flux in the [SIV] line was found by the Spitzer IRS to be 3.73 Jy. The total TEXES flux is 5.8 Jy with an absolute uncertainty of at least $\pm30\%$.   It is not very useful to compare observations of such radically different spatial and spectral resolution as  IRS and TEXES results quantitatively;  the systematic uncertainties are too high.   We can only say that the  flux in the TEXES  $\sim1.4\arcsec$ beam is consistent with the entire galaxy's emission seen by  Spitzer.  This agrees with \citet{BT02}'s picture of the II Zw 40 starburst, which found almost all the galactic  infrared and radio emission to be produced in the central 1.5\arcsec~ core.  We therefore identify the [SIV] source with the thermal radio peak and use the radio position, which has accuracy $\pm0.01\arcsec$, rather than  the 
position from offset pointing which has lower accuracy.

The medium-resolution spectra are asymmetric, with some structure on the red side of the peak. But with the low spectral resolution of only $\sim$25\kms , little more can be said about the line profile.  We accordingly observed [SIV] on 10-11 January in the high-resolution mode, which has the same spatial plate scale but 0.003\wno/pixel and effective resolution $\sim3$\kms.  The NS slit was placed on the radio position and peaked up,  and the spectrum was nodded along the slit.  The high resolution spectrum is shown in Figure 4.  

\section{The [SIV] Line Profile}

 The appearance of the medium resolution spectra is consistent across all the slit positions.  The line profile can be formally fit  very well, with reduced $\chi^2 \approx1$,  by one Gaussian of 110$\pm10$\kms~ FWHM, centered at 779$\pm7$\kms (Figure 3).  
 The residuals to this fit show a slight red excess. 

 The high resolution spectrum makes a very different impression: on inspection it has obvious structure on the red side of the line.  There is a dip in the line profile (marked on Figure 4)   
 at $\sim800$\kms~and a secondary peak at $\sim805$\kms.  
   If we fit the line with one gaussian the residuals have the distinct feature of the secondary peak. A  two-Gaussian fit to the high-resolution data puts the stronger of the two peaks at  767$\pm1.5$\kms~and the weaker at 811$\pm2.2$\kms.  The first has FWHM 51$\pm1.3$\kms and the second 37$\pm2$\kms.   The two-Gaussian fit has  reduced $\chi^2\approx 1.2$, somewhat better than a one Gaussian fit  for this data, and shows only  very weak structure in the residuals. 
 
 Are the red structure and the apparent  secondary peak real or an observational artifact?  While there are other features of contrast similar to the dip in the profile in our spectra,  they are associated with known sources of noise.  Noise increases over wavelengths of transition between echelon orders, which appear in the current data as a dip at 
 $\sim862$\kms~, just off the [SIV] line.  Noise also goes up near atmospheric features, but the atmospheric transmission over the [SIV] line is, as discussed above, uniform to a very high degree.  In addition, it is important to note that while the medium and high resolution spectra appear  different they actually agree very well.  If the high resolution spectrum is convolved with a 27 \kms ~wide Gaussian the result is very close to the medium resolution profile.  That the red structure is seen in very similar form in each of the three independant medium resolution spectra therefore supports the reality of the secondary peak.  We conclude that the dip in the profile is significant and that the secondary peak  is real.

 \subsection{Interpreting the Line Profile}
 \subsubsection{Spatial Structure in the Star Formation Region}

The star formation in II Zw 40 
is confined to   
a $0.5~kpc$~diameter \HII\ region, which has  WR features and dominates the optical emission in both continuum and nebular lines emission. 
 The infrared and radio emission associated with the starburst comes from a source within the optical emission region, but of  much smaller extent. 
   \citet{BT02} studied the infrared and free-free radio emission from the starburst source with sub-arcsecond resolution, and concluded that the starburst consists of radio-infrared supernebulae excited by embedded super star clusters. With  
 resolution of 0\farcs1-0\farcs3 they 
 localized the star formation to  
 a core of~ $1.5\arcsec~ (\approx75~pc) $ diameter containing $\approx10^4$ O stars 
consisting of two sources extended 3\arcsec\ (150~pc) in the east-west direction,
  with a total ionization equivalent to 14,000 O7 stars.   This core, shown in Figure 2, emits the bulk of the radio and infrared flux for the entire galaxy. 
 From the radio spectrum Beck et al deduced that the emission is optically thick at cm wavelengths, a sign that the emitting regions have high electron density and small size.   The 
 highest resolution radio image ($0.12\times0.14$\arcsec) show the core breaking up into 4 unresolved sub-sources; 
 owing to the limited uv coverage and signal-to noise of these high resolution maps, these 
 could be independent peaks or density peaks in an extended structure. 
  If these four sources are real, their peak fluxes  
  lead to estimated sizes of $\approx 1~pc$ and stellar populations of $\approx~600$ O stars for the unresolved emitting regions.  
  
\subsubsection{Outflow or Double Source?} 
 The line profile shows that there is a component of  gas, in bulk about 1/4 the total of the ionized gas in the starburst, which is offset in velocity redward of the main emission.  The simplest physical models that can produce such a line profile are a red outflow or a double source. 
 
  In the  
red outflow model, 
 all the gas is physically associated with and ionized by the same embedded star cluster, but the line profile is not symmetric because the gas preferentially flows away from the cluster in one direction. This can be caused by a density gradient in the embedding cloud, and the gas flow may be driven by pressure or by the winds of the embedded stars.  This is the physical situation believed to hold in many Galactic ultra-compact \HII ~regions \citep{ZH} and  deduced from the line profile of the RISN in NGC 5253 \citep{BLT12}, with the caveat that {\it in all those sources the observed outflow is blue of the main velocity.}   Blue outflow sources, which can be presumed to on the near side of the obscuring cloud, are obviously easier to observe than red outflows which must be on the far side of significant amounts of obscuring material; in their sample of Galactic \HII regions \citet{ZH} do not detect a single red outflow source.  
 
 The sign of the velocity offset in II Zw 40 is itself a strong argument that the line shape is not due to an outflow, but perhaps not definitive.  The sources known to have blue outflows are immersed in ample obscuring material and have high extinction even at infrared wavelengths and even to the near side of the emitting region. II Zw 40 is unusual among RISN for its low extinction: as mentioned above,  it has less obscuration at infrared wavelengths than any other RISN known.  
If a red outflow is to be observable anywhere it would be in II Zw 40.  But even with the stated arguments in support of an outflow model,  we believe the second model, of a double source, is the more physically likely, and we discuss that in the next section.

 \section{Two Embedded Clusters}
 
The most likely explanation of the asymmetric line profile is that it combines two distinct sources at different velocities.  The first evidence is from the line profile itself, which shows a distinct dip and rise to a second peak.     Although the dip in the line profile is not strong,  we argue (section 3 above) that it is real.  These features cannot be explained by a smooth outflow. Further evidence to support this picture comes from the map of free-free radio emission, which shows multiple sources  \citep{BT02} in the beam.   We conclude that the high excitation starburst in II Zw 40 contains two embedded clusters or groups of clusters at different velocities. The brighter, which we will call II Zw 40-1 is at velocity consistent with the galaxy's, and II Zw 40-2 is the weaker redshifted one.  

In this  two-source model of the [SIV] emission region, what can we say about the individual sources?  The line profile is consistent with each source being a simple Gaussian, with no outflow features.  From the FWHM and an assumed $T_e$ of $13,000$ (the value in the optically observed ionized gas) the non-thermal turbulent broadening $\approx 46$\kms~ for the main source and $\approx 31$ \kms\ for the secondary. If we assume $R\approx 1~pc$ for each cluster and virial velocities, the total mass in the main source is $\approx 3\times10^5 M_\odot$ and in the smaller,  $\approx 1.4\times10^5 M_\odot$.  Clusters with these total masses and Kroupa IMFs extending to $0.1M_\odot$ would have total luminosities of $10^9$ and $4.7\times10^8~L_{\odot}$ and ionizations $4-7$ and $1.9-3.2\times10^{52}~s^{-1}$, respectively.  The total infrared luminosity and ionization of II Zw 40 are $2\times10^9 L_{\odot}$ and $1\times10^{53}~s^{-1}$ in 1\arcsec ~\citep{BT01}.  The fluxes are roughly consistent with what the two embedded clusters are calculated to produce; if the small apparent excesses of observed luminosity and ionization are real, they could show the presence of some minor star formation activity outside the two star clusters.   

Note that the above derivation found the masses, luminosities and ionizations of the embedded clusters from the assumption that they are gravitationally bound, their line widths and the size estimated from the radio spectrum and maps.  That the derived quantities are so consistent with the observed is in turn a justification of the size estimate.  It argues that the clusters cannot be more than a few parsec in radius or their mass, luminosity and ionization would be unrealistically large.

\subsection{Comparison to the Radio and Optically Observed Compact Sources}
 
Multiple compact sources have been found at other wavelengths in the complex central region of II Zw 40. Can the two kinematically distinct clusters be identified with any of them?  \citet{VA08} found 2 star clusters in the central region of II Zw 40 which can be seen in the optical and near-infrared, but they point out that all the radio emission is coincident with the one cluster they call II Zw 40-A. Their source B is a few arcseconds from the main peak and would not fall into the same TEXES beam,  and so cannot be identified with II Zw 40-2.  

It is more likely that the kinematically distinct sources are related to the 
sub-sources seen in the radio maps of \citet{BT02}.  In a 2 cm map with a $0.21\arcsec \times 0.24\arcsec$ beam (reproduced in our Figure 2),  one central source with two sub-peaks appears.  At the same wavelength and with a  $0.21\arcsec \times 0.14\arcsec$ beam the central source holds two distinct sub-sources, separated by about $0.4 \arcsec$ or $20~pc$. At even higher resolution of $0.14 \arcsec \times 0.12 \arcsec$~ the sources appear to break up further, for a total of  4 statistically significant  sources strong enough, and with enough ionization, to be detectable in our spectra.   We cannot tell if the sub-sources are really independent or if they are density peaks in the flux distribution of a single source.  The entire radio core covers an  area only slightly larger than 1 pixel of the TEXES array,  which  explains why we do not detect any spatial dependance of  the line profiles.  We have at present no way of knowing which of the sub-sources belong to which velocity peak, nor can we tell if each velocity peak holds 1, 2, 3 or even more sub-clusters (as long as their velocity distributions have peaks within a few \kms ~of each other, their superimposed spectrum would be indistinguishable from a single gaussian.)

\subsection{The Velocity Offset and Galactic Kinematics}

The brighter source, II Zw 40-1, is at galactic velocity, and the other  $44\pm2$\kms~red.  This is a very large offset for two sources only pc apart on the plane of the sky. What can we deduce about their kinematics and those of II Zw 40?  

First, the clusters are unlikely to be interacting strongly with each other.  They are almost certainly not gravitationally bound to each other: with the minimum possible relative velocity of $44\pm2$\kms, they would have to be separated by $\lesssim~1.3~pc$, which is close to the deduced size of one of the sources, to be bound.   If they are at a more reasonable distance, $10~pc$ for example, their mutual gravitational potential energy would be equivalent to a velocity of $\approx 11$\kms .   
Second, the large velocity  difference between the two clusters argues that in spite of their proximity, they were probably not formed from different cloud cores 
within a single 
 molecular cloud.  Even a large molecular cloud of $100~pc$ will have internal velocity dispersion only $~20-30$ \kms \citep{So87,2004ApJ...615L..45H}.   Third, they do not agree with the velocity gradients observed in II Zw 40.
II Zw 40 is a low mass galaxy. 
 \citet{BA82}, using optical lines, see the velocities  shift by $\approx 50$ \kms~to the red over a separation of $15 \arcsec$ (750 pc), 
 much larger than the subarcsecond separation of 
 the two [SIV] clusters.   The HI observations of \citet{BK88} with a $5.9\times7$\arcsec~beam
 find large velocity gradients across the tidal tails and 
two overlapping velocity systems offset by $30-50$ \kms , which the authors attribute to two HI clouds of almost equal mass, and the HI channel maps of \citet{VZ98}, which had resolution $17\times15.4$\arcsec~, see an emission component at the velocity of the secondary peak.  But as with the optical results, the HI gradients and velocity shifts  appear over distances much larger than the probable separation between the two star clusters.  

We thus argue 
that the most plausible explanation of the velocity offset of the second cluster associates it with a  
dynamical structure resulting from the merger that created II Zw 40.  HI observations have not detected the hypothesized   
structure because of their low resolution. Galaxy mergers can lead to high velocity differences across star forming regions: the dwarf starburst He 2-10, for example, an post-merger object in a more advanced stage than is II Zw 40 \citep{HT07},  hosts two 
sites of embedded clusters separated by 27\kms ~and $\approx131~pc$ on the plane of the sky.  
Mergers are also known to be excellent sites for formation of clusters (\citet{KN03}, \citet{MU11}, and references therein).  Observations to test this model in II Zw 40 would require spectral resolution of no more than a few \kms~and spatial resolution no worse than 0.1\arcsec~ and may be possible with ALMA.

\subsection{A Possible Blue Excess}
The best formal  fit to the high resolution spectrum has increased residuals on the blue side, suggesting that there may be excess blue emission not included in the two gaussians.  Excess blue emission may, as in NGC 5253 \citep{BLT12},  show that gas is flowing out of the embedded nebula or nebulae. Such outflows, driven by thermal  pressure or by stellar winds, can be important in the evolution of embedded clusters, as they make it possible for the embedded nebulae to maintain its small size and high density for much longer times than free expansion would allow.  Unfortunately the noise on the blue side of the [SIV] line is high because of the strong atmospheric line at $\approx~ 949.45$\wno\ in that galaxy
 ($\approx~620$\kms ~heliocentric velocity), and that may affect the continuum level.  The current data are not good enough in this wavelength region to permit us to analyze the residuals quantitatively; high-resolution observations of an ionic line in a better atmospheric location are needed to confirm or disprove the existence of an outflow. 

\section{Conclusions} 
TEXES spectroscopy of the [SIV] line from II Zw 40 has shown that the [SIV] emission is spatially very concentrated; all the [SIV] seen by Spitzer for the whole galaxy is emitted from an area  only $\sim1.4$~\arcsec~ or $\sim$~70~pc across.  This core starburst region is embedded in the giant  $\sim~0.5$~kpc \citep{DB97} \HII~region which dominates the optical continuum and nebular line emission of II Zw 40.  The  luminosity and ionization of the infrared and radio emission is equivalent to about  $10^4$ O7  stars.  

The  [SIV] line profile has a double peak: the main peak is at the galaxy velocity and the weaker peak $44\pm2$\kms~ to the red.  We interpret this profile as showing that there are two embedded clusters or groups of clusters at different velocities.  The best 2 gaussian fit to the line profile finds relatively narrow FWHMs of 47 and 32\kms~for the two peaks. If the line widths reflect only thermal broadening and gravitationally bound motions we deduce masses of $\approx 3\times10^5 M_\odot$ and $\approx 1.4\times10^5 M_\odot$ for the two clusters, consistent with normal starburst evolution and the total luminosity and ionization deduced from infrared and radio imaging \citep{BT02}. 

It is not surprising that the infrared line emission in II Zw 40 includes multiple embedded clusters.    II Zw 40 is known to be a good site for star cluster formation, with many such having been born in its current episode of star formation, and \citet{BT02} showed that the radio emission from the core contained sub-structure that was unresolved at $\sim0.12$~\arcsec scales. It is however surprising that there are multiple sources emitting  the [SIV] line. 
 [SIV] is a high excitation line from an ion ($S^{+3}$) that can only be created by a very hot massive star. Those stars are short lived; from the Spitzer spectrum the clusters in II Zw 40 are expected to be less than $\approx3$~Myr old \citep{RR04}.  The clusters must be coeval to a very high degree to be in this stage at the same time. 

The biggest surprise of the  double peaked line profile is that it suggests the embedded clusters are substantially offset in velocity while within a few pc  on the plane of the sky.   We relate this phenomenon to the interaction history of II Zw 40, which has its present form as the result of the merger of two smaller dwarf galaxies.  The optically observed tidal tails or streamers of II Zw 40  cover velocity ranges similar to what is seen in the [SIV];  it may be that the interaction set up high velocity shifts over small distances right into the core of II Zw 40.  

Finally, these results illustrate the power of high resolution spectroscopy.  In II Zw 40 a component of the galaxy which cannot be resolved spatially has been discovered from its kinematic signature in the line profile. This discovery was possible because of the high resolution of TEXES and the low thermal broadening of metal ionic lines.  

\acknowledgements
We thank the anonymous referee for a thorough and helpful report. TEXES observations at the IRTF are supported by NSF AST-0607312.  This research has made use of the NASA\&IPAC Extragalactic Database (NED) which is operated by the Jet Propulsion Laboratory, Caltech, under contract with NASA.

\begin{figure}
\begin{center}
\includegraphics[scale=0.7]{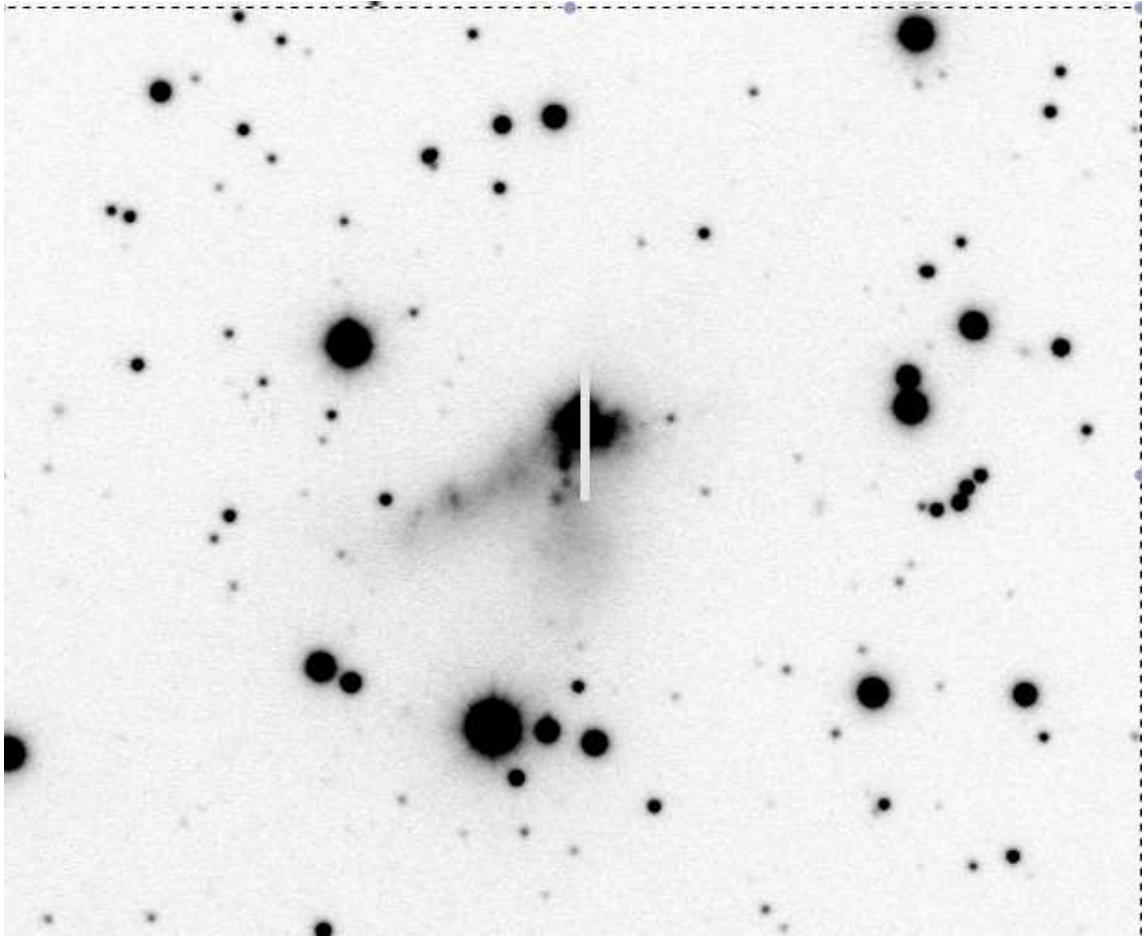}
\caption{II Zw 40 in the visible, with the high-resolution slit marked. North is up and East left. }
\end{center}
\end{figure}

\begin{figure}
\begin{center}
\includegraphics[scale=0.4]{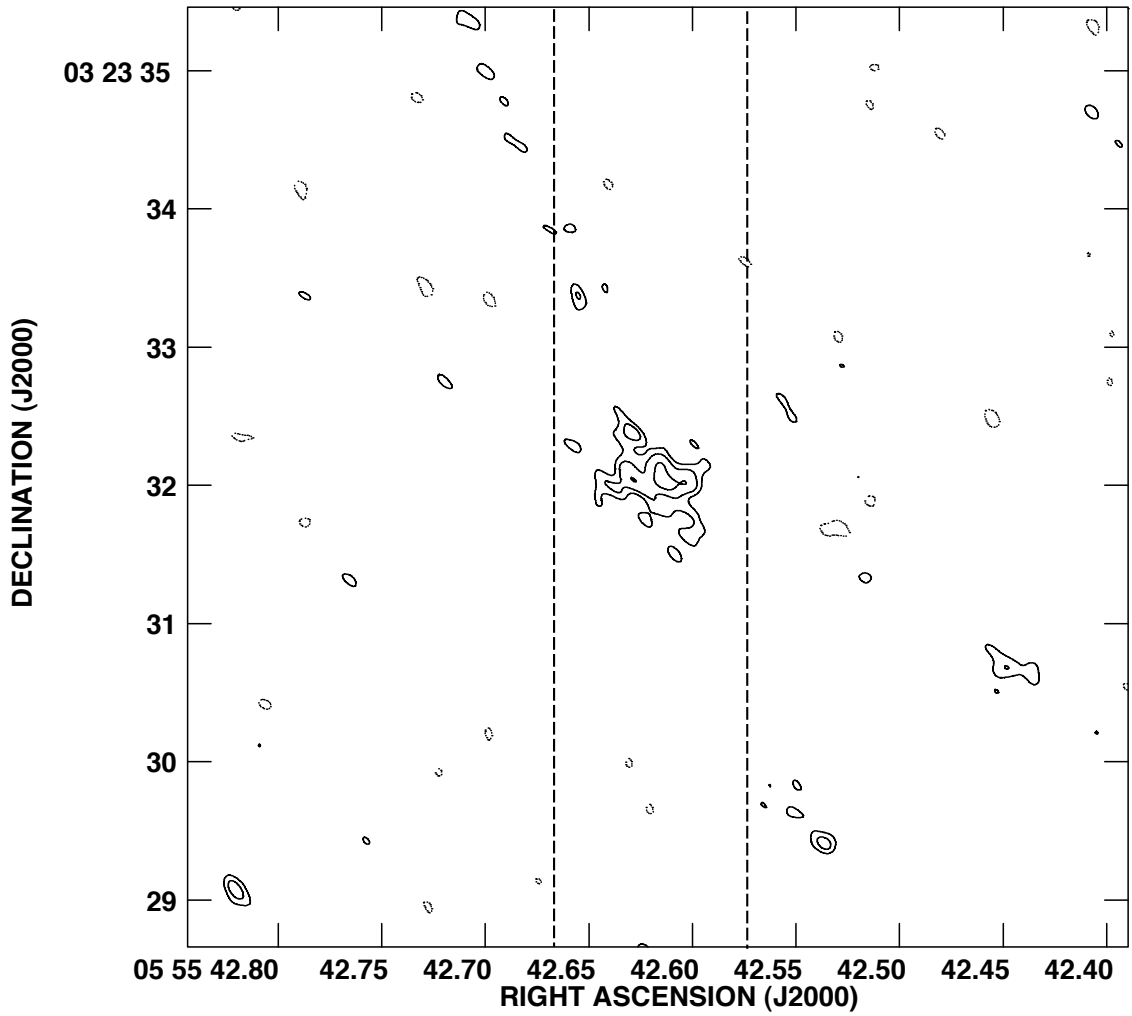}
\caption{The size of the TEXES slit, superimposed on a VLA continuum 
map of II Zw 40 at 2cm, from \citet{BT02}.  The radio contours are $\pm2^{n/2}\times 0.27$ mJy/beam and the beam size is $0.21\times 0.24$\arcsec.  The source at  $\alpha=5^h55^m42^s.69, \delta=03^o23'32.04\arcsec$ is the star formation peak and  the dominant radio and infrared source of the galaxy.  The TEXES slit illustrates the relative sizes of the infrared beam and the radio source(s) rather than the absolute position registration.  }
\end{center}
\end{figure}

\begin{figure}
\begin{center}$
\begin{array}{cc}
\includegraphics[width=3.5in, height=2.5in]{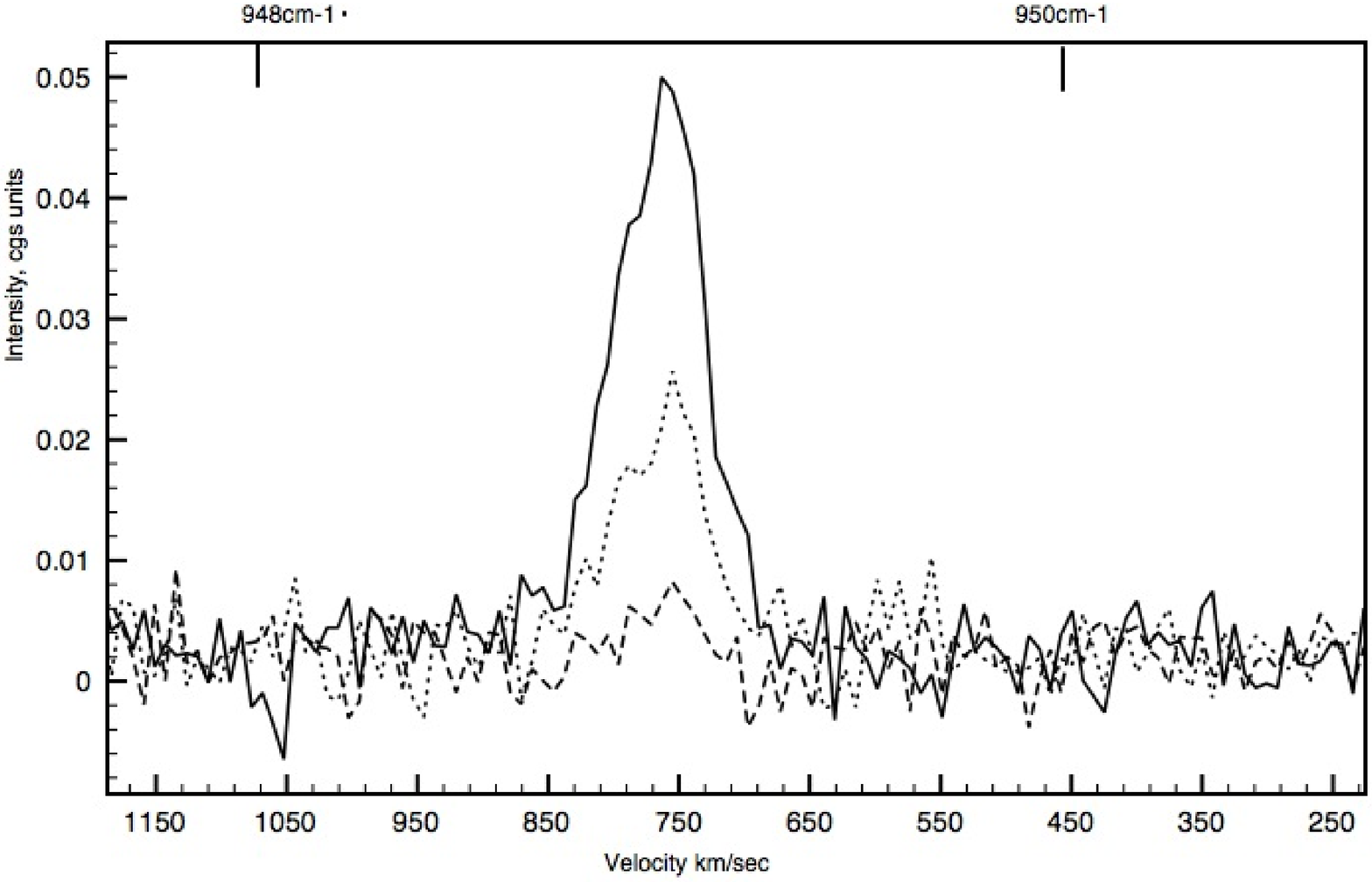} &
\includegraphics[width=3.5in]{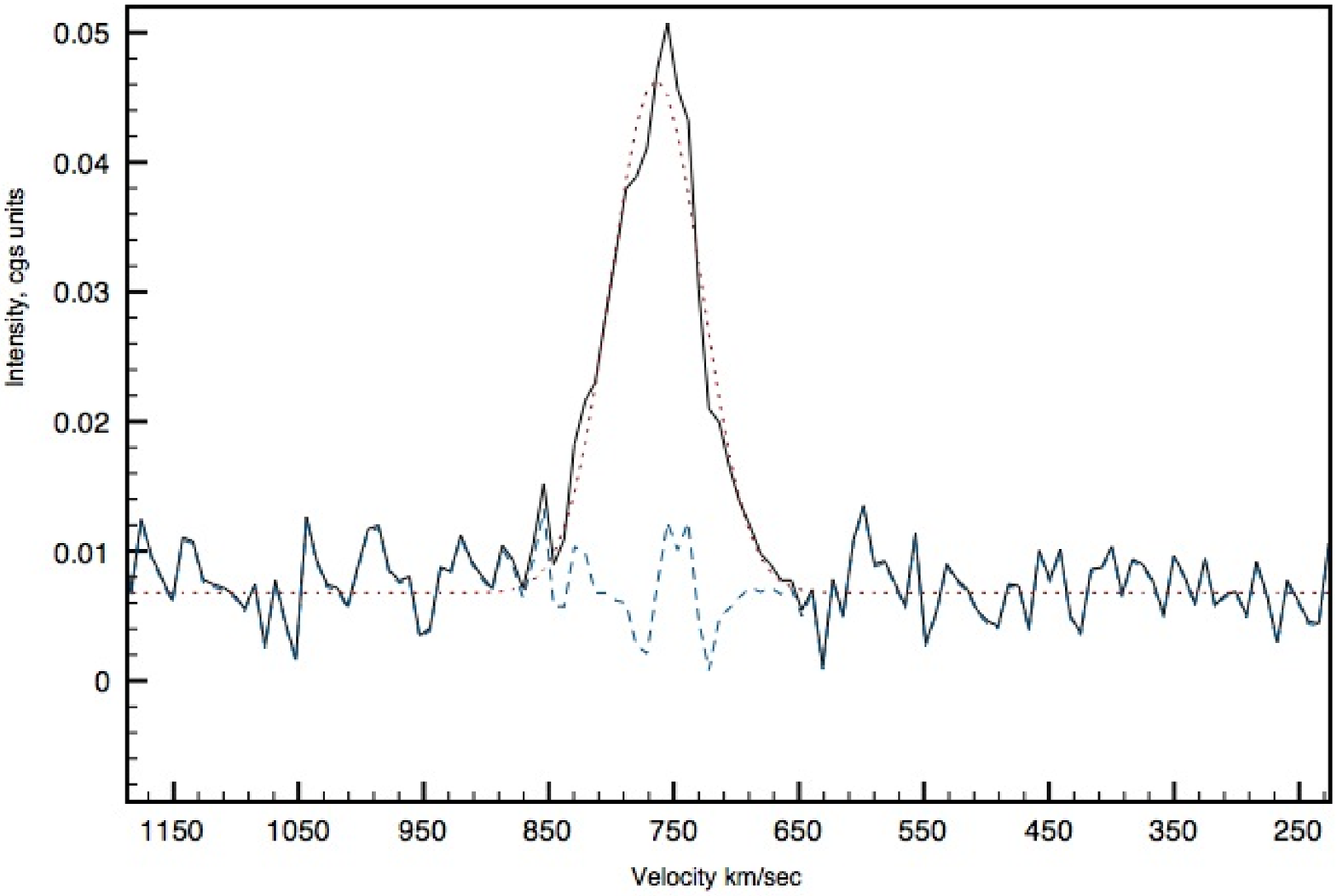}
\end{array}$
\caption{Left: the  spectrum of [S\,IV] at medium resolution in the three beam positions where it was detected. One slit was on the nominal center and the other two were offset 0.7\arcsec ~or half a slit width to east and west. The x-axis is velocity in \kms and the y-axis intensity in $ergs/s~sr~cm^2 cm^{-1}$. Two reference wavenumbers are shown at the top of the graph.  Right:   the medium resolution spectrum summed over the entire emission region and normalized to the same peak flux as Figure 2a, the best single Gaussian fit, and the residuals.}

\end{center}
\end{figure}


\begin{figure}
\begin{center}

\includegraphics[scale=0.7]{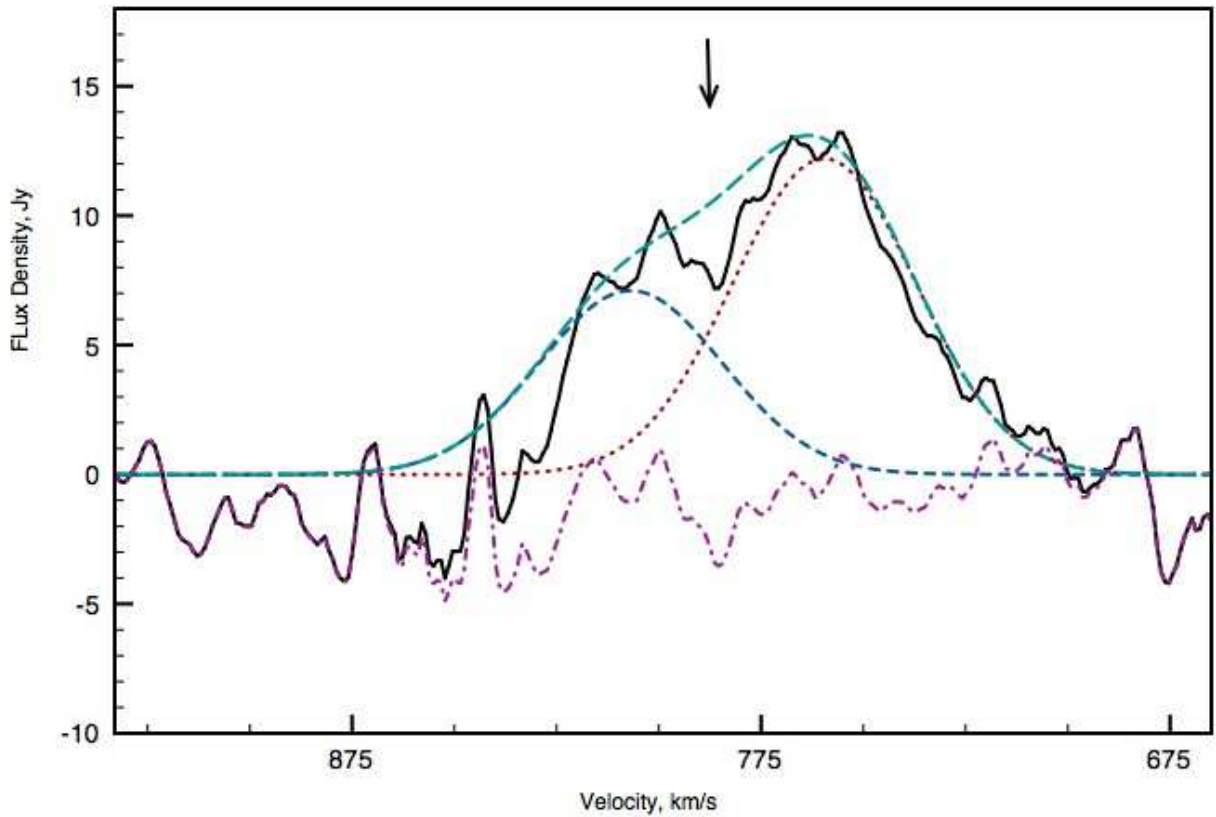}

\caption{ The high resolution spectrum, smoothed by 3 pixels or one resolution element, with the two Gaussians of the best fit, their sum,  and the residuals of the fit. The arrow indicates the "dip" in the profile. }
\end{center}
\end{figure}

\end{document}